\documentclass[twocolumn,showpacs,10pt]{revtex4-1}%
\usepackage{amsmath,amssymb,graphicx}
\usepackage{amsmath}
\usepackage{amsfonts}
\usepackage{amssymb}
\usepackage{graphicx}%
\providecommand{\U}[1]{\protect\rule{.1in}{.1in}}
\begin{document}

\title{Turbulence-immune computational ghost imaging based on a multi-scale generative adversarial network}
\author{Hao Zhang$^{1}$ and Deyang Duan$^{1,2}$}
\email{duandy2015@qfnu.edu.cn}
\affiliation{$^1$ School of Physics and Physical Engineering, Qufu Normal University, Qufu 273165, China\\$^2$Shandong Provincial Key Laboratory of Laser Polarization and Information Technology, Research Institute of Laser, Qufu Normal University, Qufu 273165, China}
\begin{abstract}
There is a consensus that turbulence-free images cannot be obtained by conventional
computational ghost imaging (CGI) because the CGI is only a classic simulation, which
does not satisfy the conditions of turbulence-free imaging. In this article, we first
report a turbulence-immune CGI method based on a multi-scale generative
adversarial network (MsGAN). Here, the conventional CGI framework is not changed,
but the conventional CGI coincidence measurement algorithm is optimized by an MsGAN.
Thus, the satisfactory turbulence-free ghost image can be reconstructed by training the
network, and the visual effect can be significantly improved.



\end{abstract}
\maketitle

Ghost imaging (GI) can be traced to the pioneering work initiated by Shih
\emph{et.al}. [1], who exploited biphotons generated via spontaneous parametric
down-conversion to realize the first entanglement-based ghost image, following
the original proposals by Klyshko [2]. The framework of quantum entangled
light source determines that the initial GI requires two light paths [1]. One of
the beams is the reference beam, which never illuminates the object and is
directly measured by a detector with spatial resolution. The other beam is
the object beam, which, after illuminating the object, is measured by a bucket
detector with no spatial resolution. By correlating the photocurrents from the
two detectors, one retrieves the ``ghost''\ image. In the debate of the
physical essence of GI [3], pseudothermal ghost imaging [4,5] and pure thermal
ghost imaging [6] have been successively confirmed, which is great
progress for GI. However, the conventional GI is not suitable for practical
applications because the structure of the two optical paths limits the
flexibility of the optical system. Fortunately, Shapiro proposed a single
optical path GI scheme: computational ghost imaging (CGI) [7,8]. Because it
has only one optical path, the imaging frame is closer to classical optical
imaging. Consequently, it has important potential applications in remote
sensing [9-11], lidar [12-16] and night vision imaging [17,18].

Atmospheric turbulence is a serious problem for satellite
remote sensing and aircraft-to-ground-based classical imaging. Surprisingly,
Meyers \emph{et. al}. found that a turbulence-free image could be obtained by
conventional dual-path GI in 2011 [19-21]. This unique and practical property
is an important milestone for optical imaging because any fluctuation index
disturbance introduced in the optical path will not affect the image quality.
Shih \emph{et. al}. revealed that the turbulence-free effect was due to the
two-photon interference [21,22]. Li \emph{et. al}. summarized the
necessary conditions for turbulence-free imaging [23]. However, CGI is a
classical simulation of GI, so it does not satisfy the conditions of turbulence-free imaging
[22]. At present, this conclusion is generally accepted. Even to this day,
it is a challenge work to realize turbulence-free CGI. Fortunately, generative adversarial network
provides a promising solution for this  [24,25].

In this article, we demonstrate that turbulence-immune CGI can be realized by a
multi-scale generative adversarial network (MsGAN). In this scheme, the basic
framework of conventional CGI is not changed. Correspondingly, we optimize
the coincidence measurement algorithm of CGI by an MsGAN. In atmospheric turbulence environment, the
satisfactory ghost images can be reconstructed by training the network. In the
following, we theoretically and experimentally illustrate this method.

We depict the scheme in Fig. 1. A quasi-monochromatic laser illuminates an
object $T\left(  \rho\right)$; then, the reflected light carrying the
object's information is received and modulated by a spatial light modulator
(SLM). A photomultiplier tube (PMT) collects the light intensity
$E_{di}\left(  \rho,t\right)  $. Correspondingly, the calculated light
$E_{ci}\left(  \rho^{^{\prime}},t\right) $ can be obtained by diffraction
theory. By processing the two light signals with a conventional CGI algorithm,
the object's image can be reconstructed, i.e.,
\begin{align}
G(\rho,\rho^{^{\prime}}) &  =\frac{1}{n}%
%TCIMACRO{\dsum \limits_{i=1}^{n}}%
%BeginExpansion
{\displaystyle\sum\limits_{i=1}^{n}}
%EndExpansion
\left(  \left\langle \left\vert E_{di}(\rho,t)\right\vert ^{2}\left\vert
E_{ci}(\rho,t)\right\vert ^{2}\right\rangle \right.  \nonumber\\
&  -\left.  \left\langle \left\vert E_{di}(\rho,t)\right\vert ^{2}%
\right\rangle \left\langle \left\vert E_{ci}(\rho,t)\right\vert ^{2}%
\right\rangle \right)  ,
\end{align}
where $\left\langle \cdot\right\rangle $ is an ensemble average. The
subscript $i=1,2,...n$ denotes the $i$th measurement, and $n$ is the total
number of measurements.

The flow chart of the MsGAN is shown in Fig. 2. The scheme mainly consists of
three parts: (i) a conventional CGI algorithm to obtain distorted images; (ii) a
generator G of MsGAN to generate random data into sample images through
continuous training; (iii) a discriminator D of MsGAN to distinguish
generated samples from real images.
\begin{figure}[ptbh]
\centering
\fbox{\includegraphics[width=1.0\linewidth]{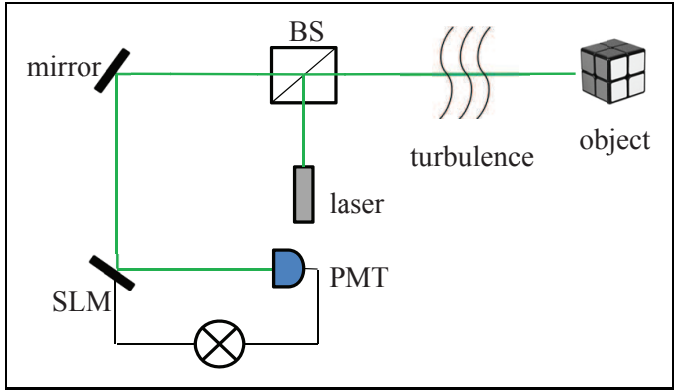}}\caption{Setup of the
computational ghost imaging. BS: beam splitter, SLM: spatial light
modulator, PMT: photomultiplier tube.}%
\label{fig:false-color}%
\end{figure}\begin{figure}[ptbh]
\centering
\fbox{\includegraphics[width=1.0\linewidth]{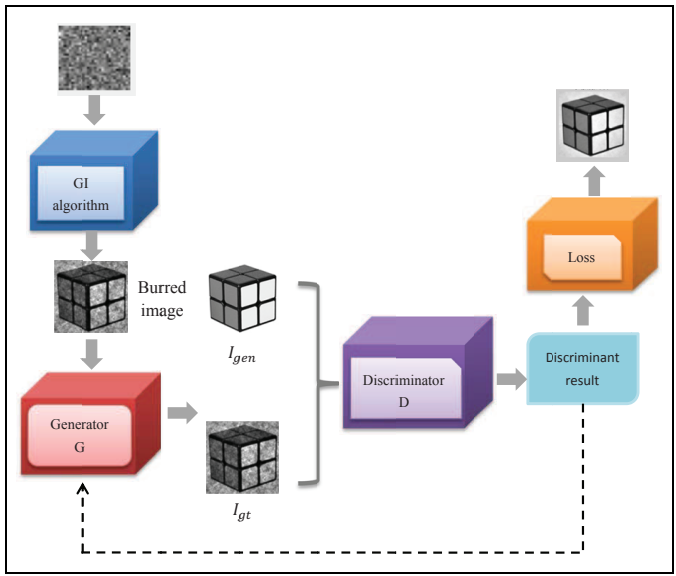}}\caption{Network
structure of the MsGAN.}%
\label{fig:false-color}%
\end{figure}

The network structure of generator G is shown in Fig. 3. The basic framework
is a symmetric U-Net architecture. On this basis, multi-scale attention
feature extraction units and a multi-level feature dynamic fusion unit are
added to the U-Net architecture. The multi-scale attention feature extraction
units use convolution kernels of different sizes to extract multi-scale
feature information in larger receptive fields. The multi-level feature fusion
unit can dynamically fuse feature images with different proportions by
adjusting the weight and excavating the semantic information of different
levels. Since U-Net will lose detailed features of the images in the process
of encoding and decoding, the visual detail features of the reconstructed
image can be improved by adding skip connections as hierarchical semantic
guidance. The network mainly consists of a full convolution part and an
up-sampling part. The full convolution part of the U-Net consists of a
pre-training convolution module and multi-scale attention feature extraction
units. The pre-trained convolution module uses the convolution layer and
max-pooling layer of the Inception-ResNet-v2 backbone network.
\begin{figure}[ptbh]
\centering
\fbox{\includegraphics[width=1.0\linewidth]{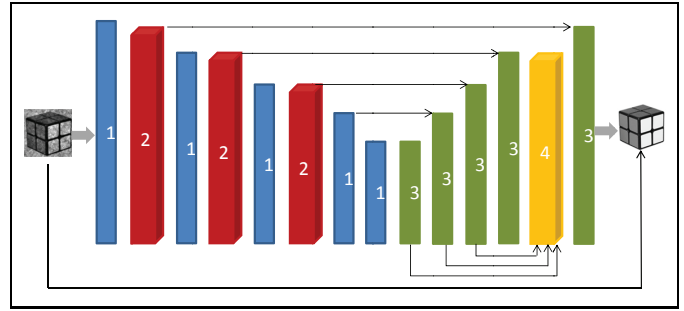}}\caption{Network
structure of the generator G. 1 represents the pre-trained convolution module;
2 represents a multi-scale attention feature extraction unit; 3 represents an
up-sampling layer and 4 represents a multi-level feature dynamic fusion unit.}%
\label{fig:false-color}%
\end{figure}

The multi-scale attention feature extraction units are composed of
multi-branch convolution layers and an attention layer (Fig. 4). The
multi-branch convolution layers are composed of different sizes of juxtaposed
dilated convolutions, which correspond to different sizes of receptive fields, to
extract various features. Three branches with receptive fields of
$3\times3$, $5\times5$, and $7\times7$ simultaneously extract features of the
input images. After the information feature graphs of different scales are
obtained, the cascaded feature graphs are readjusted to the input size through
the convolution operation. \begin{figure}[ptbh]
\centering
\fbox{\includegraphics[width=1.0\linewidth]{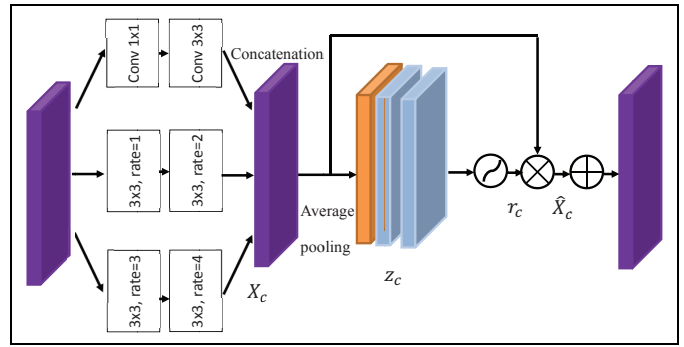}}\caption{Structure of
multi-scale attention feature extraction units.}%
\label{fig:false-color}%
\end{figure}

In feature extraction, an attention mechanism is introduced to generate
different attention for each channel feature to distinguish between
low-frequency parts (smooth or flat areas) and high-frequency parts (such as
lines, edges, and textures) of images to pay attention to and learn the key
content of the image. First, the global context information of each channel is
used to compress the spatial information of each channel by global average
pooling. The expression is%

\begin{equation}
z_{c}=G_{p}\left(  X_{c}\right)  =\frac{1}{H\times W}%
%TCIMACRO{\dsum \limits_{i=1}^{H}}%
%BeginExpansion
{\displaystyle\sum\limits_{i=1}^{H}}
%EndExpansion%
%TCIMACRO{\dsum \limits_{j=1}^{W}}%
%BeginExpansion
{\displaystyle\sum\limits_{j=1}^{W}}
%EndExpansion
X_{c}\left(  i,j\right)
\end{equation}
where $X_{c}$ is the aggregation convolution feature graph with the
size of $H\times W\times C$, and $Z_{c}$ is the compressed global pooling
layer with the size of $1\times1\times C$. ReLU and Sigmoid activation
functions are used to implement the gating principle to learn the nonlinear
synergistic effect and mutual exclusion relationship between channels, and the
attention mechanism can be expressed as%

\begin{align}
r_{c}  &  =\sigma\left\{  Conv\left[  \delta\left(  Conv\left(  z_{c}\right)
\right)  \right]  \right\}  ,\\
\overset{\symbol{94}}{X_{c}}  &  =r_{c}\times X_{c},
\end{align}
where $\delta$ and $\sigma$ are the activation functions of ReLU and sigmoid,
respectively, $r_{c}$ is the weight of excitation, $\overset
{\symbol{94}}{X_{c}}$ is the feature graph after the adjustment of the
attention mechanism. The global pooling layer $z_{c}$ goes through the
down-sampling convolutional layer and ReLU activation function, and the
channel number is recovered through the up-sampling convolutional layer.
Finally, it is activated by the sigmoid function to obtain the channel
excitation weight $r_{c}$. The value of the aggregation convolution layer
$X_{c}$ channel is multiplied by different weights to obtain the output
$\overset{\symbol{94}}{X_{c}}$ of the adaptive adjustment channel attention.

The up-sampling part of the U-Net architecture includes up-sampling layers and
a multi-level feature dynamic fusion unit. In the up-sampling part of the
generator network, the feature graphs at different levels contain different
instance information. A multi-level feature fusion unit is proposed to enhance
the information transfer among feature maps at different levels. In addition, we propose a dynamic fusion network structure to solve the problem of
feature conflict at different levels. The method assigns different weights to
the spatial position of the feature map to screen the valid features and
filter out contradictory information through learning. First, the feature
images of different scales are adjusted to the same size by up-sampling,
and spatial weights are set for feature images of different levels during
fusion to find the optimal fusion strategy. It can be specifically expressed
as: {\small
\begin{equation}
F^{\ast}=\omega_{1}\times F^{1\uparrow}+\omega_{2}\times F^{2\uparrow}%
+\omega_{3}\times F^{3\uparrow}+\omega_{4}\times F^{4\uparrow},
\end{equation}
} where $\omega_{i}$ is the weight corresponding to feature graphs of
different levels, $F^{i\uparrow}$ is the standard feature graph of the
$i$th feature graph adjusted to a uniform size after up-sampling, $F^{\ast}$
is the final feature graph output by dynamic fusion of all levels of
features through adaptive weight allocation. In the end, generator G
introduces a skip connection directly from the input to the output, which forces
the model to focus on learning residuals.

We input the feature images generated by generator G and the real images
into discriminator D for discrimination. Discriminator D adopts the
PatchGAN structure and consists of four convolution layers with $4\times4$
convolution kernels.
\begin{equation}
L_{adv}=E\left[  \left\Vert \lg\left(  D\left(  I_{gt}\right)  \right)
\right\Vert _{2}^{2}\right]  +E\left[  \left\Vert \lg\left(  D\left(
I_{gen}\right)  \right)  \right\Vert _{2}^{2}\right]  ,
\end{equation}
where $I_{gt}$ denotes real images, and $I_{gen}$ denotes generated
images, respectively. On the content loss of image reconstruction, the mean
square error loss of generated images and target images is selected to obtain
a higher peak signal-to-noise ratio (PSNR). The mean square error loss is%

\begin{equation}
L_{MSE}=\frac{1}{M\times N}%
%TCIMACRO{\dsum \limits_{i=1}^{m}}%
%BeginExpansion
{\displaystyle\sum\limits_{i=1}^{m}}
%EndExpansion%
%TCIMACRO{\dsum \limits_{j=1}^{n}}%
%BeginExpansion
{\displaystyle\sum\limits_{j=1}^{n}}
%EndExpansion
\left\Vert I_{gt,ij}-I_{gen,ij}\right\Vert ^{2},
\end{equation}
Simultaneously, to eliminate artifacts and restore
high-frequency details of images, the reconstructed images have high visual
fidelity, and the visual loss is introduced as follows%

\begin{equation}
L_{perc}=%
%TCIMACRO{\dsum \limits_{i=1}^{L}}%
%BeginExpansion
{\displaystyle\sum\limits_{i=1}^{L}}
%EndExpansion
\frac{1}{H_{i}W_{i}C_{i}}\left\Vert \phi_{i}\left(  I_{gt}\right)  -\phi
_{i}\left(  I_{gen}\right)  \right\Vert _{2}^{2}.
\end{equation}
Thus, the total loss function is defined as%

\begin{equation}
L=\alpha L_{MSE}+\beta L_{perc}+\gamma L_{adv}.
\end{equation}

The experimental setup is schematically shown in Fig. 1. A standard
monochromatic laser (30 mW, Changchun New Industries Optoelectronics Technology
Co., Ltd. MGL-III-532) with wavelength $\lambda=532$nm illuminates a $50:50$
beam splitter. The reflected light illuminates an object. The light reflected
by the object passes through the beam splitter, and the transmitted light is
modulated by a two-dimensional amplitude-only ferroelectric liquid crystal
spatial light modulator (Meadowlark Optics A512-450-850) with $512\times512$
addressable $15\mu m\times15\mu m$ pixels. Finally, a photomultiplier tube
(PMT) collects the modulated light and outputs photocurrent signals.
Correspondingly, the reference signal can be obtained through software. Atmospheric
turbulence is introduced by adding heating elements, which are from an electric furnace
placed in the optical path. Turbulence with different intensities can be
obtained by adjusting the temperature.
\begin{figure}[ptbh]
\centering
\fbox{\includegraphics[width=0.87\linewidth]{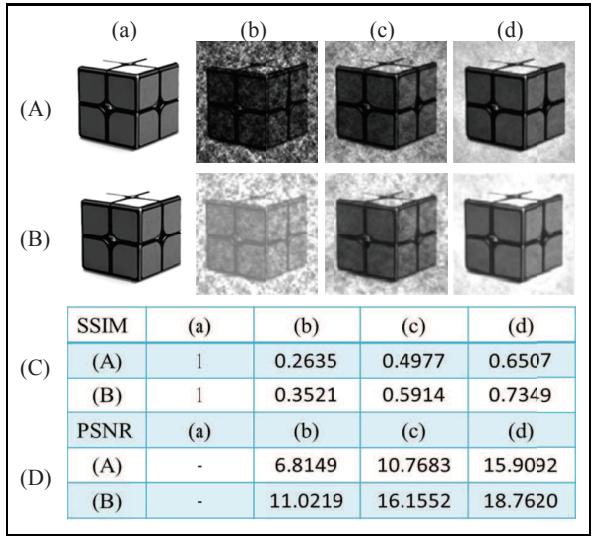}}\caption{(A) The burred
classical images caused by atmospheric turbulence, (B) Computational ghost
images reconstructed by MsGAN method, (C) and (D) The SSIM and PSNR values of
the images with different intensities of atmospheric turbulence. (a) classical
images without atmospheric turbulence, (b), (c) and (d) show the burred images
caused by different intensities of atmospheric turbulence.}%
\label{fig:false-color}%
\end{figure}

In data processing, the superparameter of the loss function is set to
$\alpha=0.5$, $\beta=0.01$, $\gamma=0.01$. Adam is used for parameter
optimization in the training process, and batch\_size is set to 1. In the
data set, 173 and 35 real object images are collected as the training set and
test set, respectively.
\begin{figure}[ptbh]
\centering
\fbox{\includegraphics[width=0.87\linewidth]{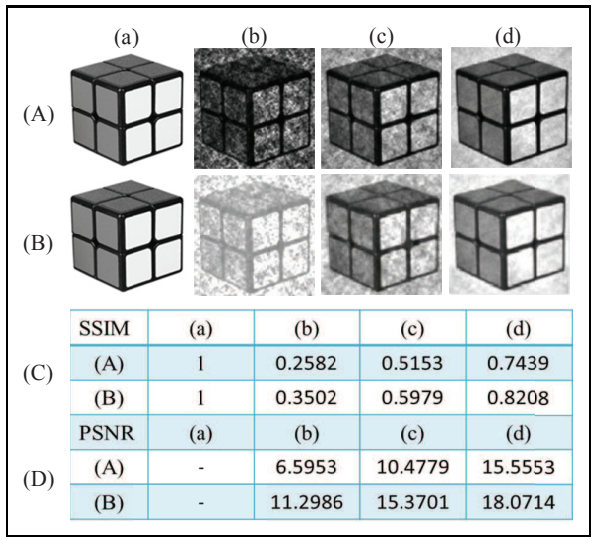}}\caption{(A) The burred
classical images caused by atmospheric turbulence, (B) Computational ghost
images reconstructed by MsGAN method. SSIM and PSNR show that this method is
effective for different objects, and the effect is almost the same.}%
\label{fig:false-color}%
\end{figure}

The effect of turbulence on a classical image is easily observed in Figs. 5A(b-d),
where three classical images were taken by a conventional CCD camera. Experimental
results show that classical images are significantly blurred by turbulence. Moreover,
with increasing turbulence intensity, the image distortion becomes more obvious.
Figs. 5B(f-h) show the experimental results of this method. From the human vision,
the reconstructed computational ghost images have better image quality than the
turbulence blurred images and are similar to the images without turbulence. To
quantitatively evaluate the effect of this method, we use the structural similarity
(SSIM) and peak signal-to-noise ratio (PSNR) to measure the image quality.
Figs. 5C and 5D show that the SSIM and PSNR of the reconstructed images are
significantly better than the burred classical images. Indeed, SSIM and PSNR
show that there is still a certain gap between the image obtained by this
method and the complete turbulence-free image. However, from the perspective
of human visual effect, the difference between the image obtained by this
method and the turbulence-free image is acceptable. Consequently, this method
is called turbulence-immune CGI. To prove the effectiveness
of this method, we choose the other Rubik's cubes as the object to experiment.
The experimental results (Fig. 6) are almost identical to above one.

In summary, the first turbulence-immune computational ghost
imaging experiment was demonstrated in this article. Although CGI framework does not satisfy the conditions of
turbulence-free imaging, the satisfactory ghost image
can be reconstructed using the MsGAN method in atmospheric turbulence environment. We hope that this
method can provide a promising solution to overcome atmospheric turbulence in
applications of CGI and single-pixel imaging.

National Natural Science Foundation of China
(11704221, 11574178, 61675115), Taishan Scholar Project of Shandong
Province (tsqn201812059).

The authors declare that there are no conflicts of interest related to this article.

Data underlying the results presented in
this paper are not publicly available at this time but may be
obtained from the authors upon reasonable request.

\end{document}